\documentclass[preprint]{aastex}

\usepackage{amssymb}
\usepackage{mathrsfs}
\usepackage{rotating}
\usepackage{graphicx}
\usepackage{epstopdf}
\usepackage{longtable,lscape}

\newcommand{\fn}[1]{\footnote{\scriptsize{#1}}} 

\newcommand{\Fig}[1]{Fig{#1}.}  
\newcommand{\etal}{{et al.}}  
\newcommand{\Cassit}{\textit{Cassini}}  
\newcommand{\Voyit}{\textit{Voyager}}  
\newcommand{\NHit}{\textit{New~Horizons}}  

\begin{document} 

\title{Observing Planetary Rings and Small Satellites with JWST:\\Science Justification and Observation Requirements\fn{Manuscript accepted by \textit{Publications of the Astronomical Society of the Pacific}}}

\author{Matthew~S.~Tiscareno$^{1,2}$\fn{Email: \texttt{matt@seti.org}}, Mark~R.~Showalter$^2$, Richard~G.~French$^3$, Joseph~A.~Burns$^1$, Jeffrey~N.~Cuzzi$^4$, Imke~de~Pater$^5$, Douglas~P.~Hamilton$^6$, Matthew~M.~Hedman$^7$, Philip~D.~Nicholson$^1$, Daniel~Tamayo$^8$, Anne~J.~Verbiscer$^9$, Stefanie~N.~Milam$^{10}$, and~John~A.~Stansberry$^{11}$}

\affil{$^1$Cornell~University, $^2$SETI~Institute, $^3$Wellesley~College, $^4$NASA~Ames~Research~Center, $^5$University~of~California~at~Berkeley, $^6$University~of~Maryland, $^7$University~of~Idaho, $^8$Canadian~Institute~for~Theoretical~Astrophysics, $^9$University~of~Virginia, \\$^{10}$NASA~Goddard~Space~Flight~Center, $^{11}$Space~Telescope~Science~Institute}

\begin{abstract}
The James Webb Space Telescope (JWST) will provide unprecedented opportunities to observe the rings and small satellites in our solar system, accomplishing three primary objectives:  1)~discovering new rings and moons, 2)~unprecedented spectroscopy, and 3)~time-domain observations.  We give details on these science objectives and describe requirements that JWST must fulfill in order to accomplish the science objectives
\end{abstract}

\section{Introduction}
The rings that adorn the four giant planets are of prime importance as accessible natural laboratories for disk processes, as clues to the origin and evolution of planetary systems, and as shapers as well as detectors of their planetary environments \citep{Ringschapter13,Hedman15}.  The retinue of small moons accompanying all known ring systems are intimately connected as both sources and products, as well as shepherds and perturbers, of the rings.  Leading sources of data on ring systems include spacecraft such as \Cassit{} and \Voyit{}, but also  space telescopes such as Hubble~(HST) and Spitzer, as well as ground-based telescopes. 

Additionally, the newly-discovered rings around the minor planet Chariklo \citep{Chariklo14} confirm for the first time that small objects and solid objects can host rings.  Due to several similarities with the known giant planet rings (e.g., orbit rate, radial structure), more detailed observations of the Chariklo rings are likely to shed light on the general workings of ring systems \citep{IceGiants14}.  Furthermore, the discovery of the Chariklo rings -- along with possible rings around Chiron \citep{Ruprecht15,Ortiz15} -- raises the question of whether rings can be observed around other minor planets also.

The James Webb Space Telescope~(JWST) is being prepared for launch in 2018 to begin a planned five-year mission.  JWST will have the capability to observe solar system objects as close as Mars (Milam \etal, this issue).  Although most of the hardware is already designed and under construction if not completed, work continues on the development of operations guidelines and software and the completion of calibration tasks.  The purpose of this paper is to identify observations of planetary rings that might be undertaken by JWST and to describe what is required for JWST to accomplish those goals. 

The three primary motivations for observing rings and small moons with JWST are 1) discovering new rings and moons, 2) unprecedented spectroscopy, and 3) time-domain observations.  Section~\ref{Obs} gives details on these science objectives.  Section~\ref{Requirements} describes requirements that JWST must fulfill in order to accomplish the science objectives, and Section~\ref{Conclusions} gives our conclusions.

\section{Observation Types \label{Obs}}

\subsection{Imaging of faint objects \label{HighRes}}

\begin{figure}[!t]
\begin{center}
\includegraphics[width=7cm]{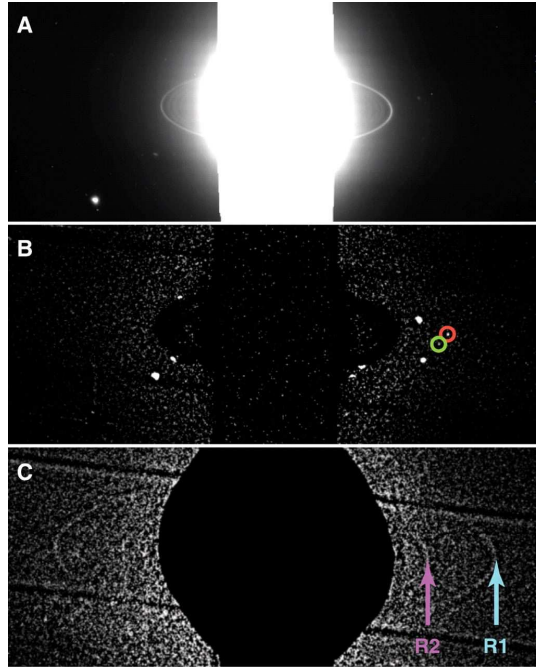}
\caption{Faint rings and moons of Uranus were discovered by Hubble in these 2003 images.  (A) Unprocessed image, (B) Filtered image showing two discoveries: Perdita (red circle), recovered 14~yr after its discovery by \Voyit{}, and Cupid (green circle), observed for the first time.  (C) Summation of 24~images showing the newly-discovered rings R1 and~R2, now respectively named the $\mu$ and $\nu$~rings.  Figure from \citet{SL06}.  \label{N14fig}}
\end{center}
\end{figure}

In the context of rings, observations of faint targets are complicated by the nearby presence of the bright planet.  Strategies are needed to enhance the apparent brightness of desired targets and/or to suppress the apparent brightness of the planet (\Fig{}~\ref{N14fig}). 

JWST will be equipped with filters (see Section~\ref{FilterQuestion}, and Milam \etal, this issue) that allow it to image giant planet systems at wavelength bands in which the planet is greatly darkened by absorption due to methane and other atmospheric constituents.  For observations of faint moons or rings that are close to bright giant planets, this will lead to greatly improved signal-to-noise and spatial resolution compared to HST and other observatories operating in the same wavelength bands (put another way, JWST will operate within the infrared methane bands at a spatial resolution comparable to that at which HST operates in visible bands, with vastly improved signal-to-noise when suppression of glare from the planet is an important factor).  

As a result, JWST will provide major advances in resolving and separating the main rings of Uranus and Neptune, improving upon HST and ground-based observations of their fine structure \citep{dePaterNeptune05,dePaterOneRing06,dePater07,SL06}.  For example, the faintest moon of Neptune, discovered using Hubble in 2013, has a V~magnitude of~26.5 \citep{ShowDPS13}, corresponding to a diameter of 16--20~km if its albedo is 0.07--0.1, as is typical for Neptune's inner moons.  While the actual scattered light contribution for JWST is not known, simple scaling would suggest that at the shortest photometric wavelength ($\sim 0.7$~$\mu$m), JWST is $\sim 2$~orders of magnitude more sensitive than Hubble.  At Neptune, this would correspond to discovering moons as small as 2~km across.  For targets that are closer and/or brighter, the size may be even smaller. 

JWST will have new sensitivity to yet-undiscovered small moons or faint rings, including the predicted rings of Mars \citep{SHN06} and Pluto \citep{StefflStern07}.  The \NHit{} spacecraft, whose flyby of Pluto will pre-date JWST, will likely not have the last word on Pluto's possible rings due to its flyby speed and limited range of viewing geometries.  JWST will be ideal for follow-up observations, possibly with greater sensitivity, and can also search for rings around other trans-Neptunian dwarf planets.  

Imaging of faint objects with JWST may be further enhanced by coronagraphy (see Section~\ref{Corona}). 

\subsection{Spectroscopy of faint objects \label{Spectro}}

The compositional diversity of solid objects in the outer solar system is apparent from the near-infrared spectra of bodies such as Triton, Pluto and Charon, which show absorption features of varying strengths due to varying amounts of methane, water and other ices on their surfaces \citep{deBergh13}. The smaller moons and rings of Neptune might have originally been made of the same stuff as these larger objects, but they also would have had much different evolutionary histories (perhaps less thermal processing, more pollution from infalling matter, etc.). Comparing the surface composition of these smaller objects to their larger neighbors should therefore help clarify the origins and histories of both, but it is difficult to obtain good-quality spectra of these very small and/or faint objects from ground-based observatories.

With its large mirror and high-quality spectrometer (Milam \etal, this issue), JWST will be able to take spectra of very faint objects.  Potential targets include the rings and small moons of Uranus and Neptune, which have never been the subjects of high-fidelity spectroscopic study, as Voyager~2 did not carry a spectrometer capable of detecting them.  Characterizing their chemical compositions is of considerable interest for addressing the origins of the Uranus and Neptune systems as well as for addressing the question of why the Uranian and Neptunian rings are so qualitatively different from those of Saturn \citep{Rochedens13}.  

\begin{figure}[!t]
\begin{center}
\includegraphics[height=3.7cm]{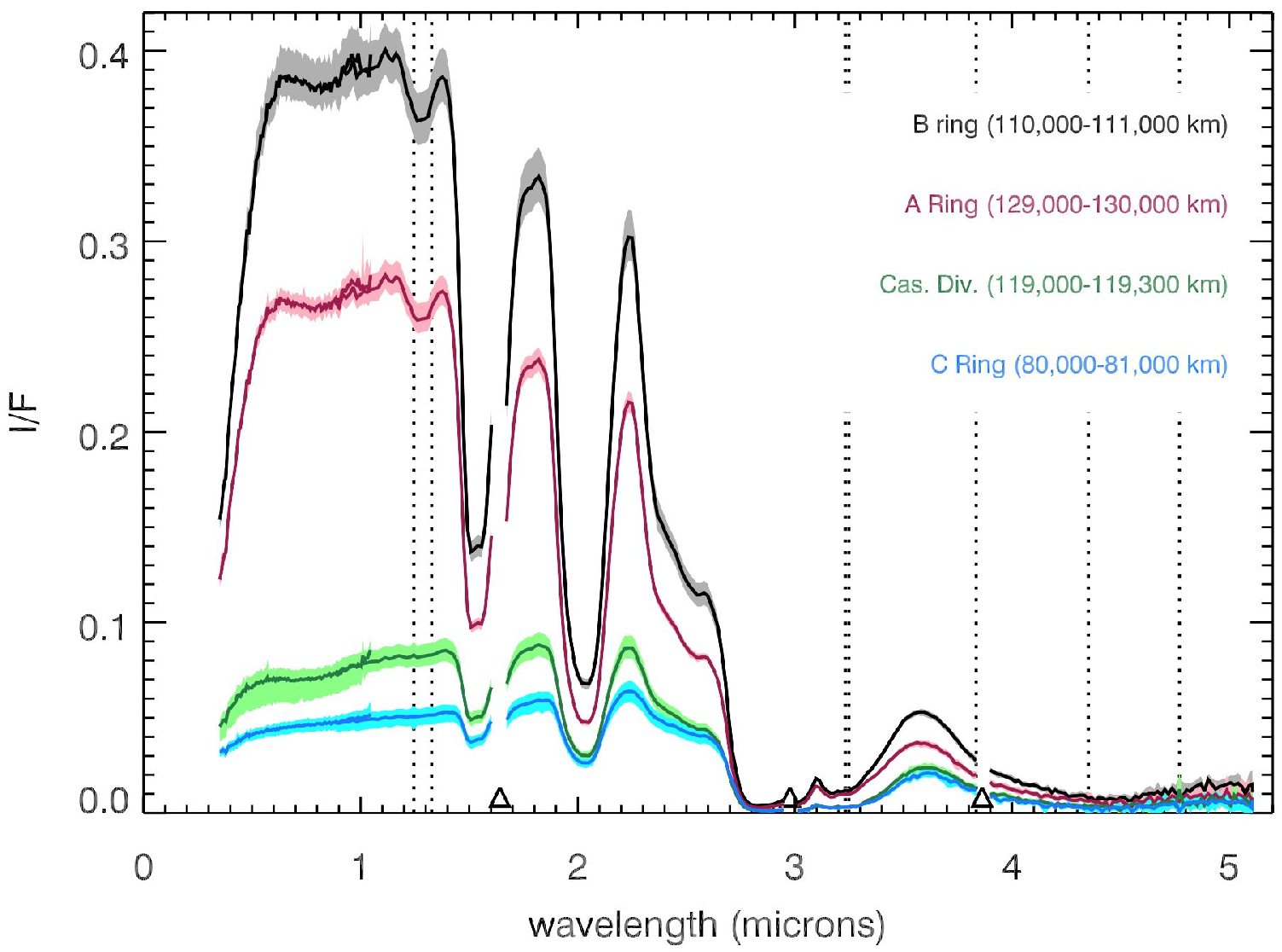}
\includegraphics[height=3.7cm]{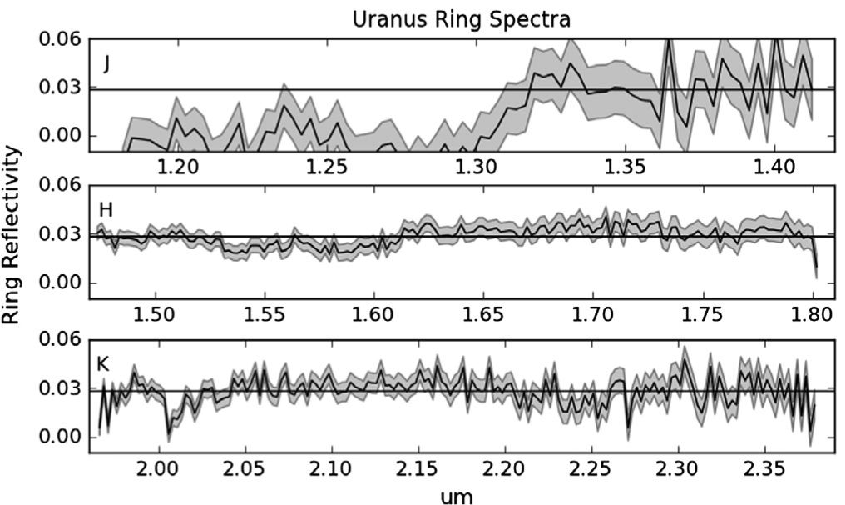}
\includegraphics[height=3.7cm]{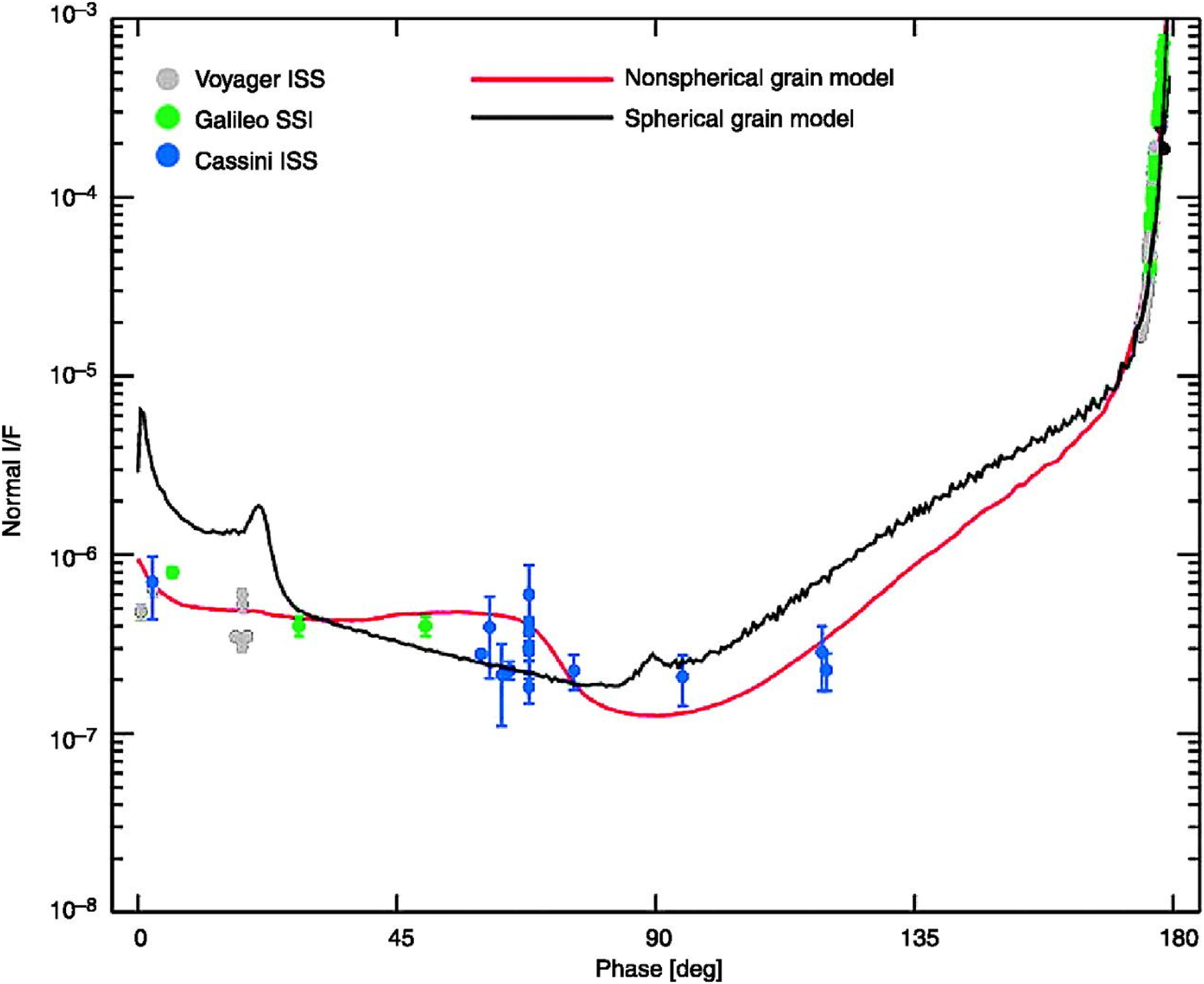}
\caption{(left) Average \Cassit{}~VIMS spectra of the lit face of selected regions in Saturn's main rings (color-coding described in panel) at low phase angles; from \citet{HedmanSpectra13}.  
(middle) Keck spectra of Uranus' main rings; from \citet{deKleer13}.  
(right) The brightness of Jupiter's main ring at visible wavelengths as a function of phase angle; from \citet{Porco03}.  
\label{RingSpectra}}
\end{center}
\end{figure}

By the same token, JWST will be able to acquire very sensitive spectra of all objects over a broad range of wavelengths.  It will be able to fill in the gap between Cassini~VIMS and Cassini~CIRS (from 5~to 8~microns) and will be able to map Saturn's rings in the 1.65-micron water absorption feature, which falls in an internal gap in VIMS' spectral coverage and is unusual in that its depth is useful for mapping temperature variations \citep{Grundy99}.  Its spatial resolution will be a few hundred km, comparable to CIRS, and its sensitivity will be greater, so it should be capable of improving current maps of Saturn's rings in the thermal infrared (though over a very limited range of phase angles) and may achieve the first detection of the faint silicate absorption features at $\gtrsim$10~microns \citep{Crovisier97,Stansberry04,Emery06}, yielding information about the little-understood non-water-ice components of Jupiter's and Saturn's rings. 

What we do know about the spectra of giant planet rings, as they are likely to be seen by JWST, is shown in \Fig{}~\ref{RingSpectra}.  Only Saturn's rings have detailed spectra at low phase angles, taken by \Cassit{}~VIMS \citep{HedmanSpectra13}.  The best spectra taken to date of Uranus' rings were taken by the Keck telescope in Hawaii \citep{deKleer13}, but they are very noisy and are ripe for improvement by JWST.  Important advances from JWST spectroscopy can also be expected for Neptune's rings and Jupiter's rings, as no spectra of quality have been taken of either system at low phase angles.  Galileo~NIMS took spectra of Jupiter's rings at very high phase angles \citep[e.g.,][]{Brooks04}, but these are dominated by diffraction and do not indicate the spectral features that JWST would see.  Also shown in \Fig{}~\ref{RingSpectra} is a phase curve for Jupiter's rings \citep{Porco03}, which indicates that the observed brightness as a ratio of the solar flux when the rings are seen face-on (that is, the normal~$I/F$) is near $5 \times 10^{-7}$ at zero phase.  Since JWST will always see Jupiter's rings nearly edge-on, the observed flux will be some $10 \times$ to $100 \times$ brighter than that.  

Spectroscopy of faint objects out to 5~microns with JWST may be further enhanced by coronagraphy (see Section~\ref{Corona}). 

\subsection{Time-domain science \label{TimeDomain}}

\begin{figure}[!b]
\begin{center}
\includegraphics[height=5.5cm]{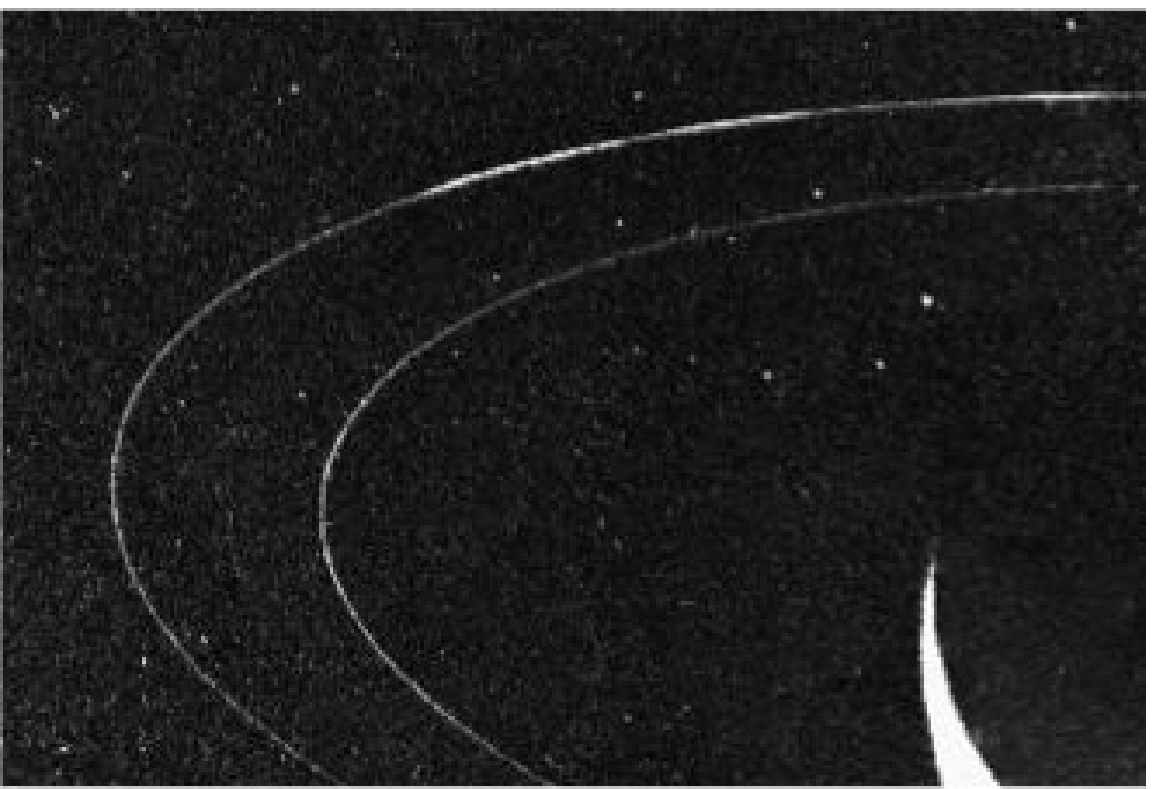}
\hspace{0.2cm}
\vspace{0.25cm}
\includegraphics[height=5.5cm]{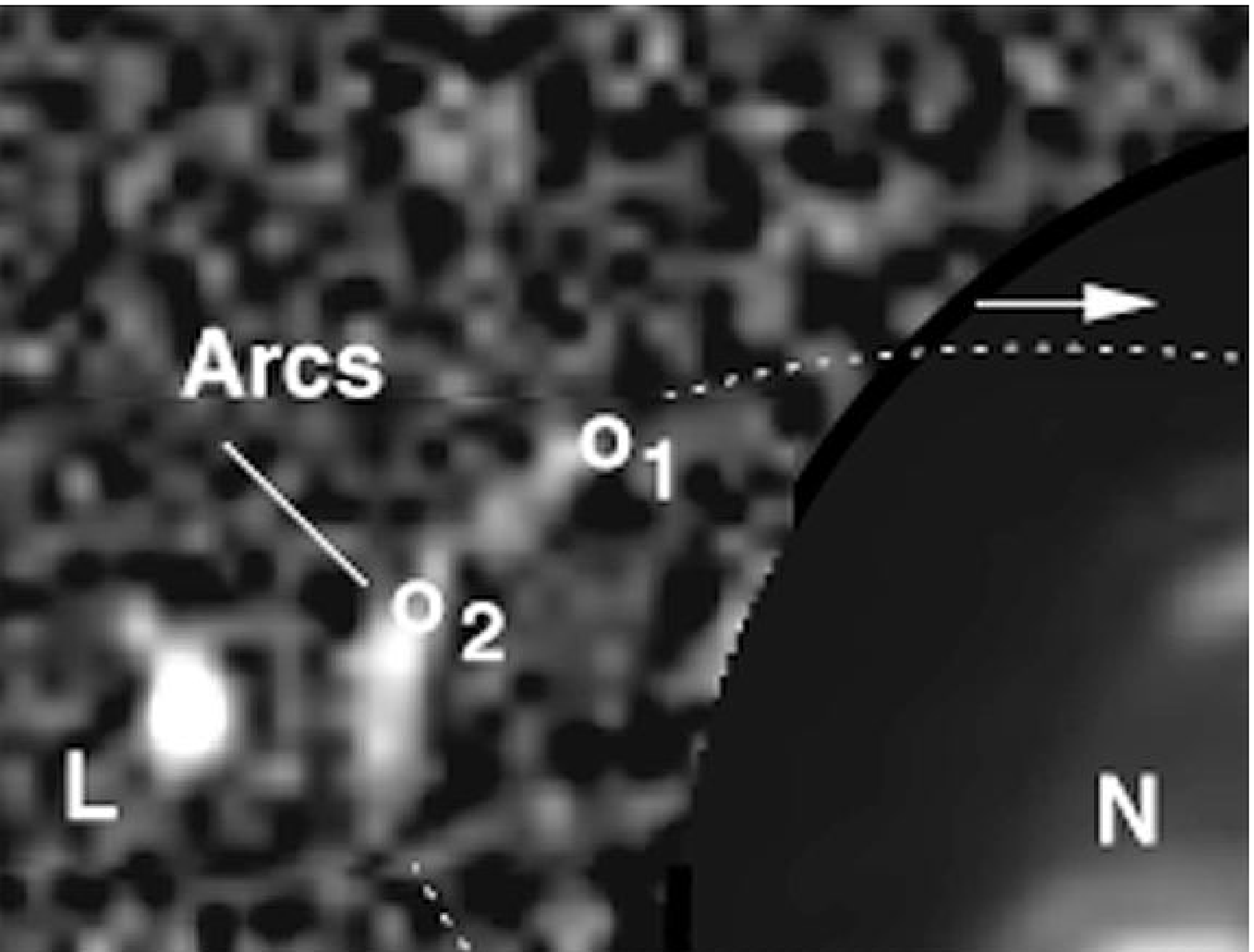}
\includegraphics[width=12cm]{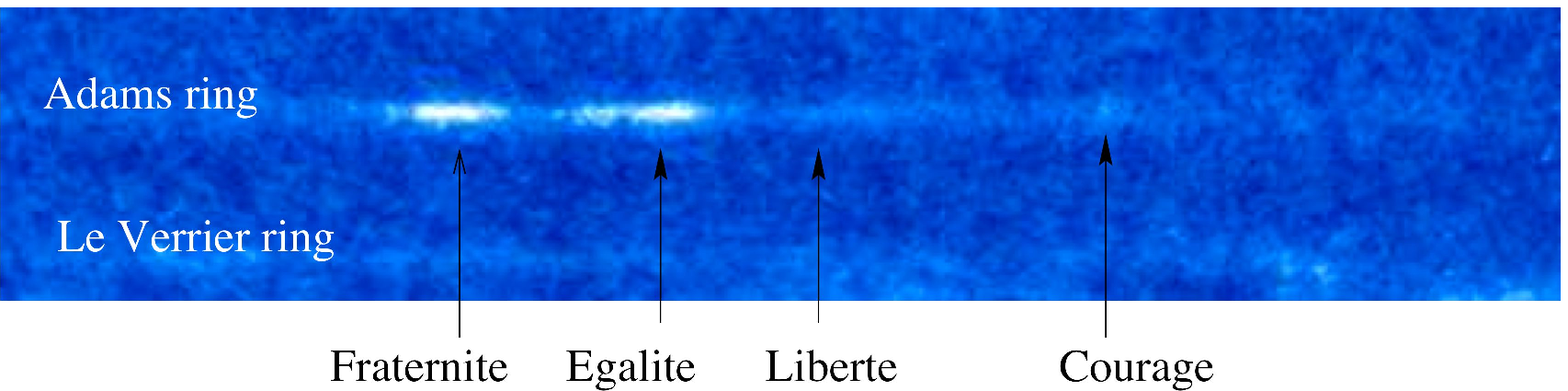}
\caption{The ring arcs of Neptune were first imaged by \Voyit{} (upper left), then reacquired in 1996 and beyond by Hubble \citep[upper right, from][]{Dumas99} and by the Keck telescope \citep[bottom, from][]{dePaterNeptune05}.  \label{HubbleSpokes}}
\end{center}
\end{figure}

Time-domain observations (sustained observation and tracking) of targets that are faint, recently discovered, or known to be changing is of high importance.  JWST observations will be important for continuing to characterize the chaotic orbits of moons including those of Pluto \citep{SH15}, Prometheus and Pandora at Saturn \citep{GR03a} and Mab at Uranus \citep{Kumar15}, as well as the evolving ring arcs of Neptune \citep{dePaterNeptune05}, progressively winding ripple patterns in the rings of Jupiter and Saturn that trace cometary impacts \citep{HedmanCorrugation11,ShowCorrugation11}, and other faint targets.  It may also be capable of tracking the azimuthal arcs or clumps in the rings of Jupiter \citep{Show07} and the ``propeller'' moons embedded in Saturn's rings \citep{Giantprops10}.  

JWST will serve as an important window on the outer solar system for imaging Targets of Opportunity such as the Jupiter impact of 2009 \citep{Hammel10} and the Saturn storm of 2010-11 \citep{Sayanagi13}.  

\subsection{Equinox}

\begin{figure}[!t]
\begin{center}
\includegraphics[width=11cm]{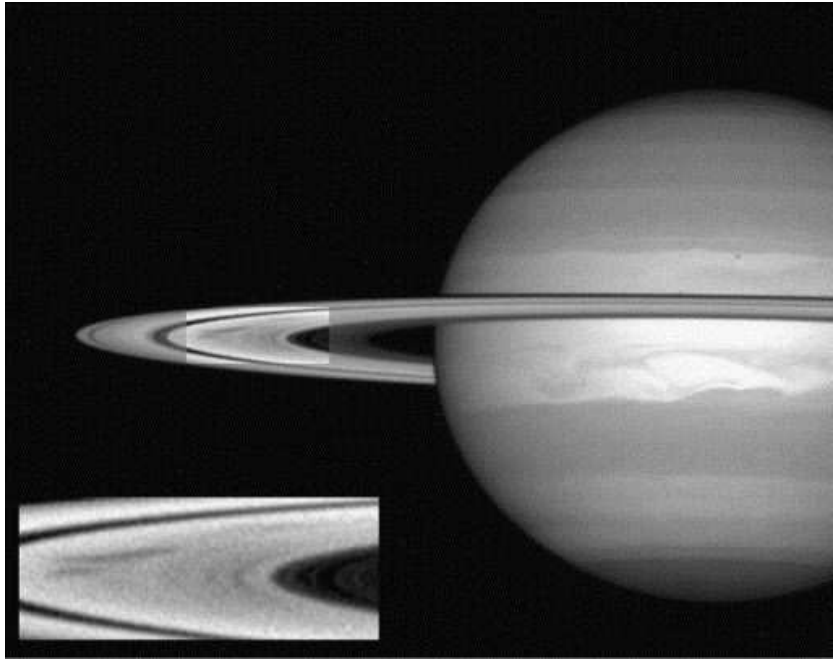}
\caption{The ghostly radial markings known as spokes, shown in the highlighted portion of this 1996 Hubble image of Saturn and its rings \citep{McGhee05}, trace the interplay of interplanetary impacts and magnetic forces in SaturnÕs rings. JWST will continue to track the seasonal behavior of spokes after the end of the Cassini spacecraft mission \citep{SpieJwst14}.  \label{HubbleSpokes}}
\end{center}
\end{figure}

The next Saturn equinox will take place in 2025.  The event itself will not be observable by JWST, as it will occur when Saturn is near the Sun as seen from Earth, but low sun angles will be observable approximately three months before and after equinox.  This will facilitate the observation of seasonal phenomena such as spokes (\Fig{}~\ref{HubbleSpokes}), which are prevalent near equinox and absent near solstice \citep{Mitchell06,Mitchell13}.  JWST will have sufficient resolution to continue monitoring spokes, as has HST \citep{McGhee05}, which will have particular value as the \Cassit{}~mission will have ended in~2017.  JWST will also be able to improve on the tracking of clumps in and around the F Ring near equinox \citep{McGhee01}, and will enjoy optimal edge-on viewing of Saturn's dusty E and G rings\fn{The Phoebe~ring, which lies in Saturn's orbit plane and is \textit{always} edge-on as seen from Earth, is thus always available for optimal edge-on viewing.} during this season \citep{dePater96eg,dePater04eg}. 

\begin{figure}[!t]
\begin{center}
\includegraphics[width=8cm]{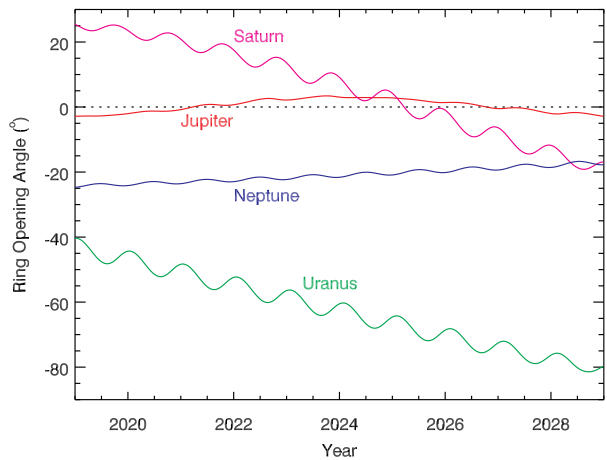}
\caption{Opening angle as a function of time for known ring systems from 2019 to 2029 \citep{Meeus97}.  The first half of this interval corresponds to the JWST prime mission.  \label{OpeningAngles}}
\end{center}
\end{figure}

Neither Uranus nor Neptune has an equinox that falls within the JWST mission (\Fig{}~\ref{OpeningAngles}).  During the JWST mission, Sun angles will decrease at Neptune 
and will increase at Uranus. 
This will lead to increasingly favorable viewing for both systems as the JWST mission progresses, since Neptune's rings are primarily dusty while Uranus' rings are dense and sharp-edged.  The only exact equinoxes possibly observable by JWST will be at Jupiter; these will provide optimal viewing of vertical structure in the halo/gossamer rings.

\subsection{Commissioning observations and Multi-instrument observation opportunities}

The rings and small satellites discipline does not currently have any clear candidates for observations to be carried out during JWST commissioning or for observations to test JWST's multi-instrument capabilities.

\section{Requirements \label{Requirements}}

\subsection{Stray light and extended point-spread functions \label{Stray}}

Because faint rings and small moons are often in close proximity to the bright planet about which they orbit, it is important to characterize the stray light and extended point-spread function (EPSF) of JWST instruments to determine how close to the planet an object can be and still be observable.  

The point-spread functions are well characterized for point sources within the field of view, but EPSFs need to be compiled from these for extended objects, especially when the planet is off the edge of the field of view.  Preliminary analysis of stray light should also be characterized through studying the blueprints of the telescope and instruments, and later by testing on the ground after the spacecraft is built.  It should be kept in mind that the severity of stray light might be different at different wavelengths.  The limited roll ability of the JWST spacecraft, which can rotate about the line-of-sight axis by only $\pm 5 ^\circ$, makes this issue more pressing, as it largely eliminates the strategy of rotating the field of view so as to spatially separate the observation target from stray light artifacts.  

\subsection{Color filters \label{FilterQuestion}}

Imaging of faint objects is enhanced at wavelength bands in which the planet is dim, especially at absorption features of atmospheric constituents such as methane (see Section~\ref{HighRes}).  The methane absorption feature at 2.3~microns is strong, and corresponds to a wavelength at which water ice is bright.  However, JWST does not have a filter centered on 2.3 microns \citep[Milam \etal, this issue]{JWST06}.  
Study is needed to determine whether imaging at 2.3~microns might be facilitated by selecting the appropriate wavelength of a NIRSpec cube using the Integral Field Unit mode \citep{JWST06}, and also whether this mode offers imaging quality as envisioned in Section~\ref{HighRes}.  This is especially germane for Jupiter and Saturn, which are too extended to provide a basis for adaptive optics, giving JWST a more pronounced advantage over ground-based telescopes.  Study is also needed to determine whether filters centered on methane's 1.8-micron and on 3.4-micron bands can offer comparable imaging quality.  

\subsection{Coronagraphy \label{Corona}}

Both imaging and spectroscopy of faint objects (Sections~\ref{HighRes} and~\ref{Spectro}) can be enhanced near a bright planet by using the NIRSpec instrument in its Microshutter Array (MSA) mode, which suppresses the brightness of the boresighted planet by a factor of $10^{4}$ while taking spectrally-resolved images.  Stray light problems (see Section~\ref{Stray}) would likely be minimized with the bright planet on the boresight, increasing the attractiveness of this technique. 

Further study is needed to characterize whether the MSA mode will facilitate good observations of faint rings or moons.  For example, is the attenuation of the planet by $10^{4}$ sufficient?  Would the angular size of Jupiter render it too large to be suppressed by this technique?  

\subsection{Stellar occultations}

Stellar occultations, in which the brightness of a star is continuously monitored as it passes behind a semi-transparent object such as a ring, can distinguish very fine structural details, often superior to the detail discernible by direct imaging albeit along only a single dimension \citep[e.g.,][]{Bosh02,Hedman07d}.  Occultations are an excellent method for determining the precession rates of rings, leading to tight and otherwise unobtainable constraints on planetary interiors \citep[e.g.,][]{Nich14c}.  Occultations require that JWST's trajectory be predicted precisely enough that potential occultations can be reliably identified ahead of time, that pointing be accurate enough to keep the star within the single-pixel field of view, and that read-out time be fast enough to maximize the spatial resolution of the data (Santos-Sanz et al., this issue).  

A minimum useful read-out time (i.e., maximum useful cadence) is set by signal-to-noise.  A high cadence will be especially important for Uranus, as its rings will be oriented roughly perpendicular to the stars' velocity as seen by JWST.  Spectrally integrated brightness measurements are adequate, though spectral resolution would yield information about the particle-size distribution \citep[e.g.][]{HedmanSpectra13}.  

In principle, occultations of smaller targets  would also be very valuable, as occultations are currently the only method available for probing structures such as Chariklo's rings \citep{Chariklo14}.  However, the small angular size of such targets makes occultations of suitably bright stars uncommon and, combined with the uncertainty in predicting JWST's trajectory, makes it difficult to predict whether occultations will occur with sufficient accuracy.

Technical details regarding the feasibility of occultations with JWST can be found in the companion paper by Santos-Sanz \etal (this issue), who find that the current JWST planning ephemeris yields excellent occultation opportunities for Saturn, Uranus, and Neptune between 2019 and 2022.

\subsection{Target acquisition, pointing accuracy, and exposure duration}

Exposure durations should be available for at least the period of time that it takes for a target moon to move by one pixel.  To increase the integration time beyond that interval, co-adding of images will be needed in any case.  The moon with the highest known apparent velocity (relative to its planet's velocity) is Metis, at 0.5 arcseconds per minute when in conjunction with Jupiter, though all moons move considerably slower on the sky near their greatest elongations.
Pointing accuracy should be finer than a few tenths of a pixel.

\section{Conclusions \label{Conclusions}}

JWST promises to be an unprecedentedly valuable observatory for observing planetary rings and their attendant small moons \citep{SpieJwst14}.  We encourage the observing community to use the ideas in Section~\ref{Obs} to formulate observation plans.  We encourage the JWST project to address the items in Section~\ref{Requirements}, to the extent possible, in order to ensure the highest possible quality of observations.  Because the post-launch servicing missions that were crucial to the success of HST will not be possible for JWST, it is imperative for both communities to identify and resolve any outstanding issues now or in the near future.


\begin{thebibliography}{}

\bibitem[{Bosh} et~al., 2002]{Bosh02}
{Bosh}, A.~S., {Olkin}, C.~B., {French}, R.~G., and {Nicholson}, P.~D. (2002).
\newblock {Saturn's F ring: Kinematics and particle sizes from stellar
  occultation studies}.
\newblock {\em \icarus}, 157:57--75.

\bibitem[{Braga-Ribas} et~al., 2014]{Chariklo14}
{Braga-Ribas}, F., {Sicardy}, B., {Ortiz}, J.~L., {Snodgrass}, C., {Roques},
  F., {Vieira-Martins}, R., {Camargo}, J.~I.~B., {Assafin}, M., {Duffard}, R.,
  {Jehin}, E., {Pollock}, J., {Leiva}, R., {Emilio}, M., {Machado}, D.~I.,
  {Colazo}, C., {Lellouch}, E., {Skottfelt}, J., {Gillon}, M., {Ligier}, N.,
  {Maquet}, L., {Benedetti-Rossi}, G., {Gomes}, A.~R., {Kervella}, P.,
  {Monteiro}, H., {Sfair}, R., {El Moutamid}, M., {Tancredi}, G., {Spagnotto},
  J., {Maury}, A., {Morales}, N., {Gil-Hutton}, R., {Roland}, S., {Ceretta},
  A., {Gu}, S.-H., {Wang}, X.-B., {Harps{\o}e}, K., {Rabus}, M., {Manfroid},
  J., {Opitom}, C., {Vanzi}, L., {Mehret}, L., {Lorenzini}, L., {Schneiter},
  E.~M., {Melia}, R., {Lecacheux}, J., {Colas}, F., {Vachier}, F., {Widemann},
  T., {Almenares}, L., {Sandness}, R.~G., {Char}, F., {Perez}, V., {Lemos}, P.,
  {Martinez}, N., {J{\o}rgensen}, U.~G., {Dominik}, M., {Roig}, F., {Reichart},
  D.~E., {Lacluyze}, A.~P., {Haislip}, J.~B., {Ivarsen}, K.~M., {Moore}, J.~P.,
  {Frank}, N.~R., and {Lambas}, D.~G. (2014).
\newblock {A ring system detected around the Centaur (10199) Chariklo}.
\newblock {\em \nat}, 508:72--75.

\bibitem[{Brooks} et~al., 2004]{Brooks04}
{Brooks}, S.~M., {Esposito}, L.~W., {Showalter}, M.~R., and {Throop}, H.~B.
  (2004).
\newblock {The size distribution of Jupiter's main ring from Galileo imaging
  and spectroscopy}.
\newblock {\em \icarus}, 170:35--57.

\bibitem[{Crovisier} et~al., 1997]{Crovisier97}
{Crovisier}, J., {Leech}, K., {Bockelee-Morvan}, D., {Brooke}, T.~Y., {Hanner},
  M.~S., {Altieri}, B., {Keller}, H.~U., and {Lellouch}, E. (1997).
\newblock {The spectrum of Comet Hale-Bopp (C/1995~01) observed with the
  Infrared Space Observatory at 2.9~AU from the Sun}.
\newblock {\em Science}, 275:1904--1907.

\bibitem[{de~Bergh} et~al., 2013]{deBergh13}
{de~Bergh}, C., {Schaller}, E.~L., {Brown}, M.~E., {Brunetto}, R.,
  {Cruikshank}, D.~P., and {Schmitt}, B. (2013).
\newblock {The ices on transneptunian objects and centaurs}.
\newblock In {Gudipati}, M.~S. and {Castillo-Rogez}, J., editors, {\em The
  Science of Solar System Ices}, pages 107--146. Springer, New~York.

\bibitem[{de Kleer} et~al., 2013]{deKleer13}
{de Kleer}, K., {de Pater}, I., {{\'A}d{\'a}mkovics}, M., and {Hammel}, H.
  (2013).
\newblock {Near-infrared spectra of the uranian ring system}.
\newblock {\em \icarus}, 226:1038--1044.

\bibitem[{de Pater} et~al., 2005]{dePaterNeptune05}
{de Pater}, I., {Gibbard}, S.~G., {Chiang}, E., {Hammel}, H.~B., {Macintosh},
  B., {Marchis}, F., {Martin}, S.~C., {Roe}, H.~G., and {Showalter}, M. (2005).
\newblock {The dynamic neptunian ring arcs: Evidence for a gradual
  disappearance of Libert{\'e} and resonant jump of Courage}.
\newblock {\em \icarus}, 174:263--272.

\bibitem[{de Pater} et~al., 2006]{dePaterOneRing06}
{de Pater}, I., {Hammel}, H.~B., {Gibbard}, S.~G., and {Showalter}, M.~R.
  (2006).
\newblock {New dust belts of Uranus: One ring, two ring, red ring, blue ring}.
\newblock {\em Science}, 312:92--94.

\bibitem[{de Pater} et~al., 2007]{dePater07}
{de Pater}, I., {Hammel}, H.~B., {Showalter}, M.~R., and {van Dam}, M.~A.
  (2007).
\newblock {The dark side of the rings of Uranus}.
\newblock {\em Science}, 317:1888--1890.

\bibitem[{de Pater} et~al., 2004]{dePater04eg}
{de Pater}, I., {Martin}, S.~C., and {Showalter}, M.~R. (2004).
\newblock {Keck near-infrared observations of Saturn's E and G rings during
  Earth's ring plane crossing in August 1995}.
\newblock {\em \icarus}, 172:446--454.

\bibitem[{de Pater} et~al., 1996]{dePater96eg}
{de Pater}, I., {Showalter}, M.~R., {Lissauer}, J.~J., and {Graham}, J.~R.
  (1996).
\newblock {Keck infrared observations of Saturn's E and G rings during Earth's
  1995 ring plane crossings}.
\newblock {\em \icarus}, 121:195--198.

\bibitem[{Dumas} et~al., 1999]{Dumas99}
{Dumas}, C., {Terrile}, R.~J., {Smith}, B.~A., {Schneider}, G., and {Becklin},
  E.~E. (1999).
\newblock {Stability of Neptune's ring arcs in question}.
\newblock {\em \nat}, 400:733--735.

\bibitem[{Emery} et~al., 2006]{Emery06}
{Emery}, J.~P., {Cruikshank}, D.~P., and {Van Cleve}, J. (2006).
\newblock {Thermal emission spectroscopy (5.2 38 {$\mu$}m) of three Trojan
  asteroids with the Spitzer Space Telescope: Detection of fine-grained
  silicates}.
\newblock {\em \icarus}, 182:496--512.

\bibitem[{Gardner} et~al., 2006]{JWST06}
{Gardner}, J.~P., {Mather}, J.~C., {Clampin}, M., {Doyon}, R., {Greenhouse},
  M.~A., {Hammel}, H.~B., {Hutchings}, J.~B., {Jakobsen}, P., {Lilly}, S.~J.,
  {Long}, K.~S., {Lunine}, J.~I., {McCaughrean}, M.~J., {Mountain}, M.,
  {Nella}, J., {Rieke}, G.~H., {Rieke}, M.~J., {Rix}, H.-W., {Smith}, E.~P.,
  {Sonneborn}, G., {Stiavelli}, M., {Stockman}, H.~S., {Windhorst}, R.~A., and
  {Wright}, G.~S. (2006).
\newblock {The James Webb Space Telescope}.
\newblock {\em \ssr}, 123:485--606.

\bibitem[{Goldreich} and {Rappaport}, 2003]{GR03a}
{Goldreich}, P. and {Rappaport}, N. (2003).
\newblock {Chaotic motions of Prometheus and Pandora}.
\newblock {\em \icarus}, 162:391--399.

\bibitem[{Grundy} et~al., 1999]{Grundy99}
{Grundy}, W.~M., {Buie}, M.~W., {Stansberry}, J.~A., {Spencer}, J.~R., and
  {Schmitt}, B. (1999).
\newblock {Near-infrared spectra of icy outer solar system surfaces: Remote
  determination of H$_{2}$O ice temperatures}.
\newblock {\em \icarus}, 142:536--549.

\bibitem[{Hammel} et~al., 2010]{Hammel10}
{Hammel}, H.~B., {Wong}, M.~H., {Clarke}, J.~T., {de Pater}, I., {Fletcher},
  L.~N., {Hueso}, R., {Noll}, K., {Orton}, G.~S., {P{\'e}rez-Hoyos}, S.,
  {S{\'a}nchez-Lavega}, A., {Simon-Miller}, A.~A., and {Yanamandra-Fisher},
  P.~A. (2010).
\newblock {Jupiter after the 2009 Impact: Hubble Space Telescope imaging of the
  impact-generated debris and its temporal evolution}.
\newblock {\em \apjl}, 715:L150--L154.

\bibitem[{Hedman}, 2015]{Hedman15}
{Hedman}, M.~M. (2015).
\newblock {Why are dense planetary rings only found between 8~AU and 20~AU?}
\newblock {\em \apjl}, 801:L33.

\bibitem[{Hedman} et~al., 2011]{HedmanCorrugation11}
{Hedman}, M.~M., {Burns}, J.~A., {Evans}, M.~W., {Tiscareno}, M.~S., and
  {Porco}, C.~C. (2011).
\newblock {Saturn's curiously corrugated C~ring}.
\newblock {\em Science}, 332:708--711.

\bibitem[{Hedman} et~al., 2007]{Hedman07d}
{Hedman}, M.~M., {Burns}, J.~A., {Showalter}, M.~R., {Porco}, C.~C.,
  {Nicholson}, P.~D., {Bosh}, A.~S., {Tiscareno}, M.~S., {Brown}, R.~H.,
  {Buratti}, B.~J., {Baines}, K.~H., and {Clark}, R. (2007).
\newblock {Saturn's dynamic D ring}.
\newblock {\em \icarus}, 188:89--107.

\bibitem[{Hedman} et~al., 2013]{HedmanSpectra13}
{Hedman}, M.~M., {Nicholson}, P.~D., {Cuzzi}, J.~N., {Clark}, R.~N.,
  {Filacchione}, G., {Capaccioni}, F., and {Ciarniello}, M. (2013).
\newblock {Connections between spectra and structure in Saturn's main rings
  based on Cassini VIMS data}.
\newblock {\em \icarus}, 223:105--130.

\bibitem[{Kumar} et~al., 2015]{Kumar15}
{Kumar}, K., {de Pater}, I., and {Showalter}, M.~R. (2015).
\newblock {Mab's orbital motion explained}.
\newblock {\em \icarus}, 254:102--121.

\bibitem[{McGhee} et~al., 2005]{McGhee05}
{McGhee}, C.~A., {French}, R.~G., {Dones}, L., {Cuzzi}, J.~N., {Salo}, H.~J.,
  and {Danos}, R. (2005).
\newblock {HST observations of spokes in Saturn's B ring}.
\newblock {\em \icarus}, 173:508--521.

\bibitem[{McGhee} et~al., 2001]{McGhee01}
{McGhee}, C.~A., {Nicholson}, P.~D., {French}, R.~G., and {Hall}, K.~J. (2001).
\newblock {HST observations of saturnian satellites during the 1995 ring plane
  crossings}.
\newblock {\em \icarus}, 152:282--315.

\bibitem[{Meeus}, 1997]{Meeus97}
{Meeus}, J. (1997).
\newblock {Equinoxes and solstices on Uranus and Neptune}.
\newblock {\em Journal of the British Astronomical Association}, 107:332.

\bibitem[{Mitchell} et~al., 2006]{Mitchell06}
{Mitchell}, C.~J., {Hor{\'a}nyi}, M., {Havnes}, O., and {Porco}, C.~C. (2006).
\newblock {Saturn's spokes: Lost and found}.
\newblock {\em Science}, 311:1587--1589.

\bibitem[{Mitchell} et~al., 2013]{Mitchell13}
{Mitchell}, C.~J., {Porco}, C.~C., {Dones}, H.~L., and {Spitale}, J.~N. (2013).
\newblock {The behavior of spokes in Saturn's B~ring}.
\newblock {\em \icarus}, 225:446--474.

\bibitem[{Nicholson} et~al., 2014]{Nich14c}
{Nicholson}, P.~D., {French}, R.~G., {McGhee-French}, C.~A., {Hedman}, M.~M.,
  {Marouf}, E.~A., {Colwell}, J.~E., {Lonergan}, K., and {Sepersky}, T. (2014).
\newblock {Noncircular features in Saturn's rings II: The C~ring}.
\newblock {\em \icarus}, 241:373--396.

\bibitem[{Ortiz} et~al., 2015]{Ortiz15}
{Ortiz}, J.~L., {Duffard}, R., {Pinilla-Alonso}, N., {Alvarez-Candal}, A.,
  {Santos-Sanz}, P., {Morales}, N., {Fern{\'a}ndez-Valenzuela}, E., {Licandro},
  J., {Campo Bagatin}, A., and {Thirouin}, A. (2015).
\newblock {Possible ring material around centaur (2060) Chiron}.
\newblock {\em \aap}, 576:A18.

\bibitem[{Porco} et~al., 2003]{Porco03}
{Porco}, C.~C., {West}, R.~A., {McEwen}, A., {Del Genio}, A.~D., {Ingersoll},
  A.~P., {Thomas}, P., {Squyres}, S., {Dones}, L., {Murray}, C.~D., {Johnson},
  T.~V., {Burns}, J.~A., {Brahic}, A., {Neukum}, G., {Veverka}, J., {Barbara},
  J.~M., {Denk}, T., {Evans}, M., {Ferrier}, J.~J., {Geissler}, P.,
  {Helfenstein}, P., {Roatsch}, T., {Throop}, H., {Tiscareno}, M., and
  {Vasavada}, A.~R. (2003).
\newblock {Cassini imaging of Jupiter's atmosphere, satellites, and rings}.
\newblock {\em Science}, 299:1541--1547.

\bibitem[{Ruprecht} et~al., 2015]{Ruprecht15}
{Ruprecht}, J.~D., {Bosh}, A.~S., {Person}, M.~J., {Bianco}, F.~B., {Fulton},
  B.~J., {Gulbis}, A.~A.~S., {Bus}, S.~J., and {Zangari}, A.~M. (2015).
\newblock {29 November 2011 stellar occultation by 2060 Chiron: Symmetric
  jet-like features}.
\newblock {\em \icarus}, 252:271--276.

\bibitem[{Sayanagi} et~al., 2013]{Sayanagi13}
{Sayanagi}, K.~M., {Dyudina}, U.~A., {Ewald}, S.~P., {Fischer}, G.,
  {Ingersoll}, A.~P., {Kurth}, W.~S., {Muro}, G.~D., {Porco}, C.~C., and
  {West}, R.~A. (2013).
\newblock {Dynamics of Saturn's great storm of 2010-2011 from Cassini ISS and
  RPWS}.
\newblock {\em \icarus}, 223:460--478.

\bibitem[{Showalter} et~al., 2007]{Show07}
{Showalter}, M.~R., {Cheng}, A.~F., {Weaver}, H.~A., {Stern}, S.~A., {Spencer},
  J.~R., {Throop}, H.~B., {Birath}, E.~M., {Rose}, D., and {Moore}, J.~M.
  (2007).
\newblock {Clump detections and limits on moons in Jupiter's ring system}.
\newblock {\em Science}, 318:232--234.

\bibitem[{Showalter} et~al., 2013]{ShowDPS13}
{Showalter}, M.~R., {de Pater}, I., {French}, R.~S., and {Lissauer}, J.~J.
  (2013).
\newblock {The Neptune system revisited: New results on moons and rings from
  the Hubble Space Telescope}.
\newblock {\em AAS Division for Planetary Sciences Meeting Abstracts},
  45:206.01.

\bibitem[{Showalter} and {Hamilton}, 2015]{SH15}
{Showalter}, M.~R. and {Hamilton}, D.~P. (2015).
\newblock {Resonant interactions and chaotic rotation of Pluto's small moons}.
\newblock {\em \nat}, 522:45--49.

\bibitem[{Showalter} et~al., 2006]{SHN06}
{Showalter}, M.~R., {Hamilton}, D.~P., and {Nicholson}, P.~D. (2006).
\newblock {A deep search for Martian dust rings and inner moons using the
  Hubble Space Telescope}.
\newblock {\em \planss}, 54:844--854.

\bibitem[{Showalter} et~al., 2011]{ShowCorrugation11}
{Showalter}, M.~R., {Hedman}, M.~M., and {Burns}, J.~A. (2011).
\newblock {The impact of Comet Shoemaker-Levy~9 sends ripples through the rings
  of Jupiter}.
\newblock {\em Science}, 332:711--713.

\bibitem[{Showalter} and {Lissauer}, 2006]{SL06}
{Showalter}, M.~R. and {Lissauer}, J.~J. (2006).
\newblock {The second ring-moon system of Uranus: Discovery and dynamics}.
\newblock {\em Science}, 311:973--977.

\bibitem[{Stansberry} et~al., 2004]{Stansberry04}
{Stansberry}, J.~A., {Van Cleve}, J., {Reach}, W.~T., {Cruikshank}, D.~P.,
  {Emery}, J.~P., {Fernandez}, Y.~R., {Meadows}, V.~S., {Su}, K.~Y.~L.,
  {Misselt}, K., {Rieke}, G.~H., {Young}, E.~T., {Werner}, M.~W.,
  {Engelbracht}, C.~W., {Gordon}, K.~D., {Hines}, D.~C., {Kelly}, D.~M.,
  {Morrison}, J.~E., and {Muzerolle}, J. (2004).
\newblock {Spitzer Observations of the Dust Coma and Nucleus of
  29P/Schwassmann-Wachmann 1}.
\newblock {\em \apjs}, 154:463--468.

\bibitem[{Steffl} and {Stern}, 2007]{StefflStern07}
{Steffl}, A.~J. and {Stern}, S.~A. (2007).
\newblock {First constraints on rings in the Pluto system}.
\newblock {\em \aj}, 133:1485--1489.

\bibitem[{Tiscareno}, 2013]{Ringschapter13}
{Tiscareno}, M.~S. (2013).
\newblock {Planetary rings}.
\newblock In {Oswalt}, T.~D., {French}, L., and {Kalas}, P., editors, {\em
  Planets, Stars, and Stellar Systems, Volume~3: Solar and Stellar Planetary
  Systems}, pages 309--376. Springer, Dordrecht.
\newblock (arXiv:1112.3305).

\bibitem[{Tiscareno}, 2014a]{SpieJwst14}
{Tiscareno}, M.~S. (2014a).
\newblock {James Webb Space Telescope's astounding view of the solar system}.
\newblock {\em SPIE Newsroom}.
\newblock (11~April~2014), doi: 10.1117/2.1201404.005406.

\bibitem[{Tiscareno}, 2014b]{IceGiants14}
{Tiscareno}, M.~S. (2014b).
\newblock {Reassessing narrow rings at Uranus and Neptune}.
\newblock {\em Workshop on the Study of the Ice Giant Planets}.
\newblock (http://www.hou.usra.edu/meetings/icegiants2014).

\bibitem[{Tiscareno} et~al., 2010]{Giantprops10}
{Tiscareno}, M.~S., {Burns}, J.~A., {Srem{\v c}evi{\'c}}, M., {Beurle}, K.,
  {Hedman}, M.~M., {Cooper}, N.~J., {Milano}, A.~J., {Evans}, M.~W., {Porco},
  C.~C., {Spitale}, J.~N., and {Weiss}, J.~W. (2010).
\newblock {Physical characteristics and non-keplerian orbital motion of
  ``propeller'' moons embedded in Saturn's rings}.
\newblock {\em \apjl}, 718:L92--L96.

\bibitem[{Tiscareno} et~al., 2013]{Rochedens13}
{Tiscareno}, M.~S., {Hedman}, M.~M., {Burns}, J.~A., and {Castillo-Rogez},
  J.~C. (2013).
\newblock {Compositions and origins of outer planet systems: Insights from the
  Roche critical density}.
\newblock {\em \apjl}, 765:L28.

\end{thebibliography}
\end{document}